\documentstyle[epsfig,12pt,aaspp4]{article}
\lefthead{Naik et al.}
\righthead{KS~1947+300 with {\it BeppoSAX}}
\slugcomment{Accepted for publication in The Astrophysical Journal}
\newcommand\sax{{\it BeppoSAX}}
\newcommand\gro{KS~1947+300}
                                                                                                    
\received{}
\begin{document}
\title {Broad band X-ray spectrum of  KS~1947+300 with \sax}
\author{S. Naik\altaffilmark{1,2}, P. J. Callanan\altaffilmark{2}, B. Paul\altaffilmark{3}, and T. Dotani\altaffilmark{1}}
\altaffiltext{1}{Institute of Space and Astronautical Science, 3-1-1 Yoshinodai, Sagamihara, Kanagawa 229-8510, Japan,~naik@astro.isas.jaxa.jp, dotani@astro.isas.jaxa.jp}
\altaffiltext{2}{Department of Physics, University College Cork, Cork, Ireland,~paulc@miranda.ucc.ie}
\altaffiltext{3}{Tata Institute of Fundamental Research, Homi Bhabha Road, Mumbai 400 005, India,~bpaul@tifr.res.in}

\begin {abstract}
We present results obtained from three \sax\ observations of the 
accretion-powered transient X-ray pulsar \gro\ carried out during the 
declining phase of its 2000 November -- 2001 June outburst. A detailed 
spectral study of \gro\ across a wide X-ray band (0.1--100.0 keV) is 
attempted for the first time here. Timing analysis of the data clearly 
shows a 18.7 s pulsation in the X-ray light curves in the above energy band. 
The pulse profile of KS~1947+300 is characterized by a broad peak with 
sharp rise followed by a narrow dip. The dip in the pulse profile shows 
a very strong energy dependence. Broad-band pulse-phase-averaged spectroscopy 
obtained with three of the \sax\ instruments shows that the energy spectrum in 
the 0.1--100 keV energy band has three components, a Comptonized component,
a $\sim$ 0.6 keV blackbody component, and a narrow and weak iron emission line 
at 6.7 keV with a low column density of material in the line of sight. We 
place an upper limit on the equivalent width of the iron $K_\alpha$ line at 
6.4 keV of $\sim$ 13 eV (for a width of 100 eV). Assuming a spherical 
blackbody emitting region and the distance of the source to be 10 kpc, the 
radius of the emitting region is found to be in the range of 14--22 km, which 
rules out the inner accretion disk as the soft X-ray emitting region. 
\end{abstract}


\section{Introduction}
Accretion powered X-ray pulsars are binary systems consisting of a 
neutron star and a stellar companion. The X-ray luminosity of these
sources depends on the rate of mass accretion from the binary companion,
and hence generally variable. Transient behavior is also observed
in many X-ray binary pulsars, most of which have a Be star companion
with an eccentric orbit (Bildsten et al. 1997). The mass donor in 
these Be binary system is a B star which is still on the main 
sequence and lying well inside its Roche surface. The energy spectra 
of the X-ray pulsars are generally described by a power-law continuum 
with a high energy cut-off. A fluorescent iron emission line at 6.4 keV 
has been observed in many X-ray pulsars which serves as a probe of the 
surrounding matter. As most of the X-ray pulsars are located in the 
Galactic plane with large interstellar absorption (Bildsten et al. 1997), 
they do not show the presence of a soft excess in the energy spectra, in 
contrast to those pulsars in the Small and Large Magellanic Clouds where 
the interstellar absorption is one to two orders of magnitude less. A 
systematic study of a sample of X-ray pulsars which show soft excess 
revealed that the soft excess is a common feature in the X-ray pulsar 
spectra and the possible origins of the soft excess are emission from 
accretion column, emission by a collisionally energized cloud, reprocessing 
by a diffuse cloud, and reprocessing by optically thick material in the 
accretion disk (Hickox et al. 2004). As broad-band X-ray spectroscopy has 
rarely been performed on transient X-ray binary pulsars, it is not 
clear whether the soft component is a common feature among these sources.

KS~1947+300 is an accretion powered X-ray pulsar consisting of a neutron 
star and a high mass stellar companion. The source was discovered on 1989 
June 08 by the TTM coded-mask X-ray spectrometer aboard the $Kvant$ module of 
the $Mir$ orbiting space station (Borozdin et al. 1990). The source 
was again detected thrice in 1989 June and July but invisible in a
1989 August observation. The transient Be X-ray binary pulsar 
GRO~J1948+32 was discovered with the Burst and Transient Source 
Experiment (BATSE) detectors on board the Compton Gamma Ray Observatory 
(CGRO) satellite in 1994 April 6 (Finger et al. 1994). The pulsar was 
detected when the 20--75 keV flux increased from about $\sim$ 25 mCrab 
to 50 mCrab: it decayed to below the detection threshold of the BATSE 
detectors within 25 days. During the outburst, 18.7 s pulsations were 
detected in the hard X-ray extending up to 75 keV. The 20--120 keV BATSE 
photon spectrum was described by a power-law spectral model with a photon 
index of $\gamma$ = 2.65 $\pm$ 0.15. $RXTE$ observations of KS~1947+300 
during an outburst in 2000 October revealed pulsations of 18.76 s which 
is consistent with the transient pulsar being identical with GRO~J1948+32 
having slowed down from the 1994 April spin period of 18.70 s at an average 
rate of 8 ms yr$^{-1}$ (Swank \& Morgan 2000). This made the identity of 
KS~1947+300 with GRO~J1948+32 virtually certain. 18.70 s pulsations were
also detected in the \sax\ observations of the pulsar during the declining
phase of the 2000-2001 outburst (Naik et al. 2006). 

A search through 5 years of RXTE/ASM data  provided evidence for 41.7 day 
orbital periodicity of \gro\ (Levine \& Corbet 2000). A joint fit to the 
data obtained from a series of RXTE observations of \gro\ during the 2000-2001 
outburst and a smaller outburst in 2002 July, along with the BATSE measurement 
from the 1994 outburst, lead to an accurate measurement of the orbital 
parameters. The orbit is very close to being circular, quite unexpected 
for such an wide orbit with $a_x$ sin$i$= 137 lt-s. Assuming the mass of 
the neutron star in the binary system to be 1.4 M$_\odot$, and using the 
derived orbital parameters, Galloway et al. (2004) estimated the lower 
limit on the mass of the companion to be 3.4 M$_\odot$.
The absence of eclipses in the ASM or RXTE-PCA light curves 
rules out an inclination $i$ $\geq$ 85$^\circ$. In one occasion during the 
2000-2001 outburst, Galloway et al. (2004) found an increase in pulse 
frequency by $\sim$ 1.8 $\times$ 10$^{-6}$ Hz in $\leq$ 10 hr over and 
above the mean trend without any indication of any large increase in 
X-ray flux. The luminosity dependence of the pulse profiles of the pulsar
is found from the occasional presence of the pulsar in the field of view
of the INTEGRAL observations of the Galactic plane (Tsygankov \& Lutovinov 
2005). Using the magnetized neutron star model, they have estimated the
distance and magnetic field of the source to be 9.5$\pm$1.1 kpc and 
2.5 $\times$ 10$^{13}$ G respectively.

The optical counterpart to the transient X-ray source \gro\ 
is a moderately reddened $V$ = 14.2 early type Be star 
located at an approximate distance of $\sim$ 10 kpc in an area of 
low interstellar absorption slightly above the Galactic plane 
(Negueruela et al. 2003). Assuming the intrinsic luminosity of 
the star to be normal for its spectral type, the X-ray luminosity 
during the peak of the 2000 October outburst reveals that it was  
a typical Type~II outburst of a Be/X-ray transient. However, although 
the source has been observed several times with the BATSE and $RXTE$, the 
spectral properties of the pulsar at low energies are not well understood. 
In this paper, we present the timing and spectral properties of \gro\ 
using data from the Low Energy Concentrator Spectrometer (LECS), the 
Medium Energy Concentrator Spectrometers (MECS), and the hard X-ray 
Phoswich Detection System (PDS) instruments of the \sax\ in the energy 
band of 0.1--100.0 keV during the decaying phase of a major outburst in 
2000 November -- 2001 June. We compare these properties of \gro\ with 
those of other accreting binary X-ray pulsars.

\begin{figure}
\vskip 5.6 cm
\includegraphics{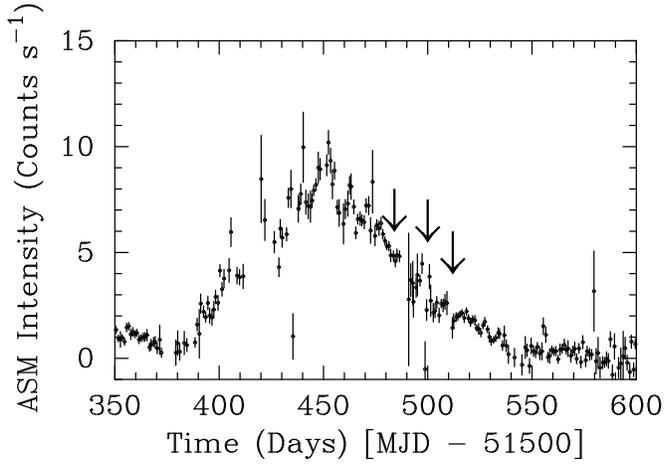}
\caption{The RXTE-ASM light curve of \gro\ in 1.5-12 keV energy band, from 
2000 September 23 (MJD 51800) to 2001 May 31 (MJD 52060). The arrows mark 
the dates of the \sax\ observations which are used for the analysis.}
\label{asm}
\end{figure}

\section{Observations}

Several X-ray outbursts of \gro\ have been observed with the $RXTE/ASM$, 
including a prominent outburst in 2000 November -- 2001 June followed by a 
few outbursts of lower intensity. During the decline phase of the 2000 
November -- 2001 June outburst, the pulsar was observed with the \sax\ 
narrow field instruments on 2001 March 16, 2001 April 1 and 2001 April 13. 
The $RXTE/ASM$ light curve of the source between 2000 September 23 and 2001 
May 31 is shown in Figure~\ref{asm}. The arrow marks in the figure indicate 
the \sax\ observations of the pulsar. The details of the \sax\ observations 
are given in Table~\ref{obs}. We have used data from the LECS, MECS and PDS 
instruments on-board \sax\ satellite. The LECS, MECS, and PDS detectors are
sensitive in 0.1--10.0 keV, 1.3--10.0 keV, and 15.0-300 keV respectively. 
The MECS consists of two grazing incidence telescopes with imaging gas 
scintillation proportional counters in their focal planes. The LECS uses an 
identical concentrator system as the MECS, but utilizes an ultra-thin entrance 
window and a driftless configuration to extend the low-energy response to 0.1 keV. 
The PDS detector is composed of 4 actively shielded NaI(Tl)/CsI(Na) phoswich 
scintillators with a total geometric area of 795 cm$^2$ and a field of view of 
1.3$^o$ (FWHM). The time resolution of the instruments during these observations 
was 15.25 $\mu$s and energy resolutions of LECS, MECS, and PDS are 25\% at 0.6 
keV, 8\% at 6 keV and $\leq$ 15\% at 60 keV respectively. The effective area of 
LECS at 0.28 keV is 22 cm$^2$ and that of MECS detector at 6.4 keV is 150 cm$^2$. 
For a detailed description of the \sax\  mission, we refer to Boella et al. (1997) 
and Frontera et al. (1997).

\begin{figure}
\vskip 7.5 cm
\includegraphics{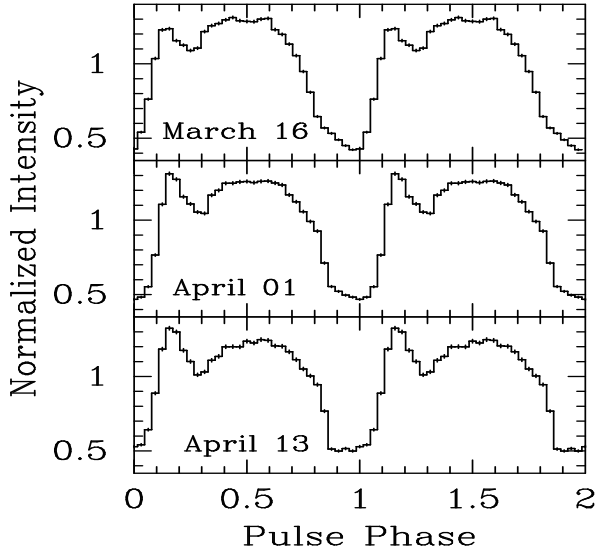}
\caption{The MECS pulse profiles in 1.3-10.0 keV energy band, obtained by 
using corresponding pulse periods, of \gro\ are shown for the three \sax\ 
observations; 2001 March observation at the top panel, 2001 April 01 
observation at the middle panel, and 2001 April 13 observation at the 
bottom panel. Two pulses are shown for clarity. The epoch is adjusted for 
each observation to obtain the minimum at phase zero. The errors are estimated
for 1 sigma confidence level.}\label{efold_all}
\end{figure}

\begin{table}
\centering
\caption{\sax\ observations of \gro}
\begin{tabular}{llllll}
\hline
\hline
Year of        &Start Time  &End Time     &\multicolumn{3}{c}{Exposure (ks)}\\
Obs.    &(Date, UT)  &(Date, UT)   &LECS         &MECS      &PDS\\
\hline
\hline
2001 March          &16, 18:12  &17, 15:22  &5    &31    &15\\
2001 April          &01, 05:58  &02, 05:07  &13   &39.5  &18\\
2001 April          &13, 01:50  &14, 01:15  &15.5 &40    &18.5\\
\hline
\hline
\end{tabular}
\label{obs}
\end{table}

\section{Timing Analysis}

We have used data from LECS, MECS, and PDS detectors for timing analysis.
The arrival times of the photons collected in all detectors were first
converted to those appropriate for the solar system barycenter. Light curves 
with a time resolution of 0.1 s were extracted from the LECS and MECS event 
data using circular regions of radius $4 \arcmin$ and $6 \arcmin$ around the
source. Light curves in the energy band of 15-150 keV with same time
resolution were extracted from the PDS data for all three \sax\ observations.
For the measurement of the pulse period, pulse folding and a $\chi^2$
maximization method was applied to all the light curves. We have derived
the pulse periods of \gro\ to be 18.70696(7) s, 18.70641(6) s, and
18.70969(5) s on 2001 March 16, 2001 April 01 and 2001 April 13
respectively. The quoted uncertainties (3$\sigma$ confidence level) 
in the pulse periods represent the trial periods at which the $\chi^2$ 
decreases from the peak value by three standard deviation of the $\chi^2$ 
values over a wide period range far from the peak. The pulse period 
measurements have been corrected for the Doppler shift due to the 
orbital motion of the neutron star using orbital parameters determined
by Galloway et al. (2004).

\begin{figure*}
\vskip 12.0 cm
\includegraphics{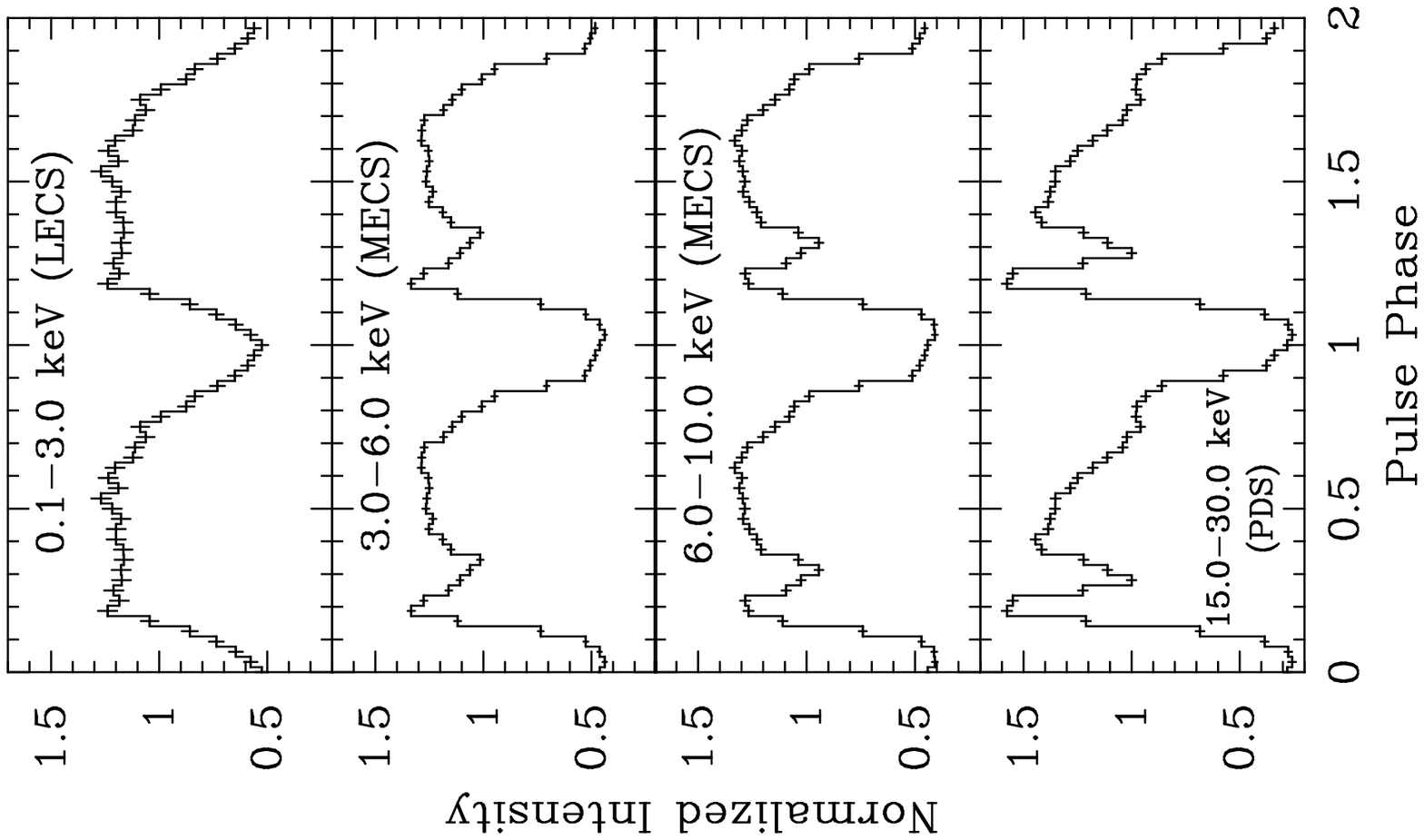}
\includegraphics{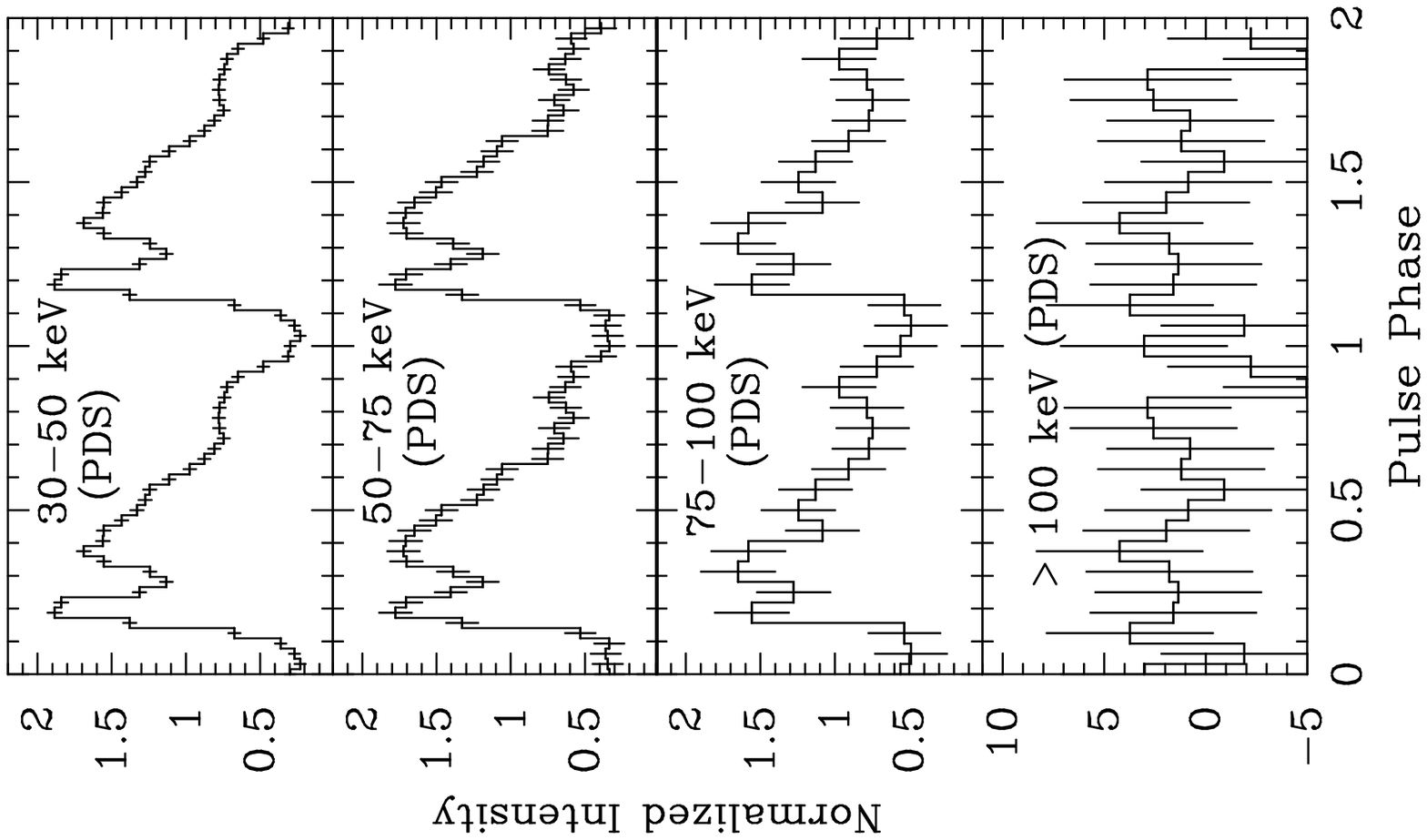}
\caption{The LECS, MECS, and PDS pulse profiles of \gro\ during the declining
phase of the 2000 December - 2001 April outburst (2001 April 01 \sax\ observation) 
are shown here for different energy bands with 32 phase bins per pulse respectively. 
The errors are estimated for 1 sigma confidence level. Two pulses in each panel are 
shown for clarity.}
\label{PP}
\end{figure*}

The pulse profiles obtained from the MECS data of the three \sax\ 
observations are shown in Figure~\ref{efold_all} with 1 sigma errors. 
It is observed that the shape of the pulse profiles in the low energy 
band (0.1--10 keV) is different to that obtained from the $BATSE$ 
observations in 20--75 keV energy band (Chakrabarty et al. 1995). The 
presence of a sharp peak at the rising part of the MECS profile followed 
by a dip like structure (Figure~\ref{efold_all}) is absent in the 20--75 
keV profile obtained from $BATSE$. To study the energy dependence of the 
pulse profiles in \gro, we generated light curves in 8 different energy 
bands from the LECS, MECS and PDS event data of the 2001 April 01 observation. 
The pulse profiles obtained from these light curves are shown in 
Figure~\ref{PP} with 1 sigma errors. It is found that the pulsations 
are detected in the light curves up to $\sim$ 100 keV, as seen in GX~1+4 
(Naik et al. 2005). In the low energy band (0.1--3.0 keV of the LECS, 
top panel), the pulse profile is flat at the top and characterized by 
the absence of the narrow peak whereas the peak is very prominent in the 
3.0--75.0 keV energy range. The peak disappears again in the 75.0--100 
keV energy band. The light curve above 100 keV is mainly background 
dominated and pulsations are not detected above 100 keV.

\begin{figure*}
\vskip 5.3 cm
\includegraphics{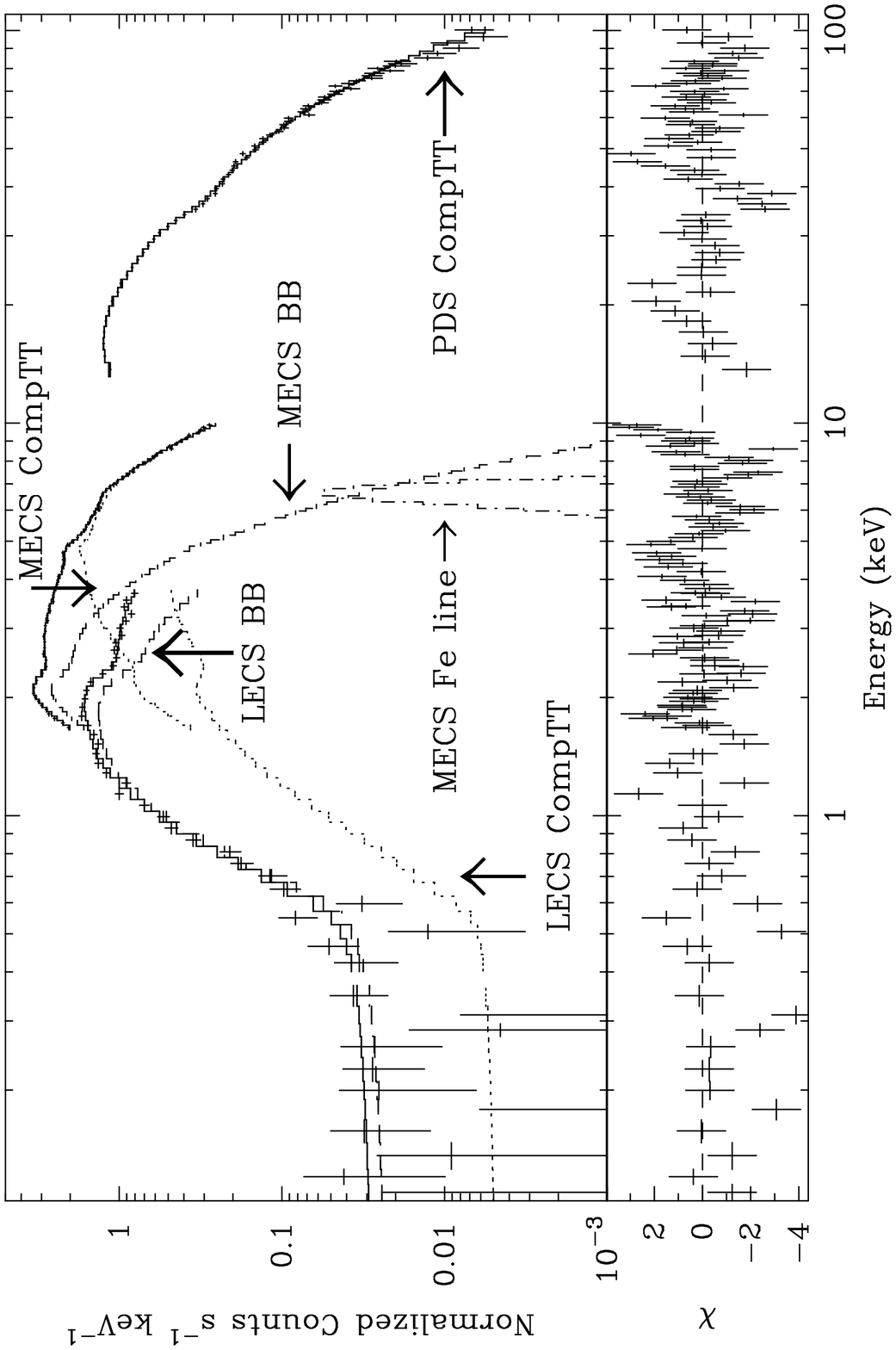}
\includegraphics{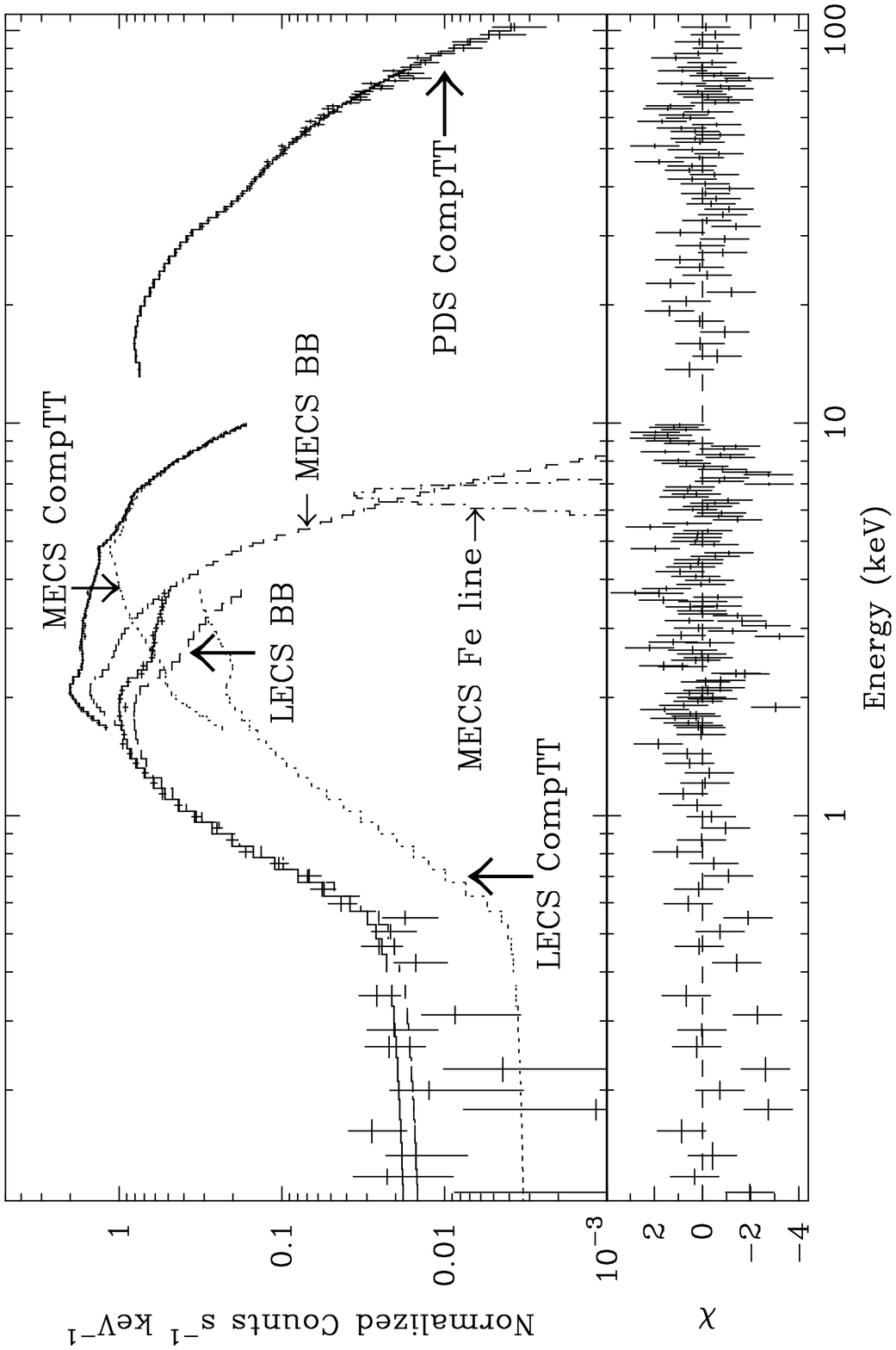}
\includegraphics{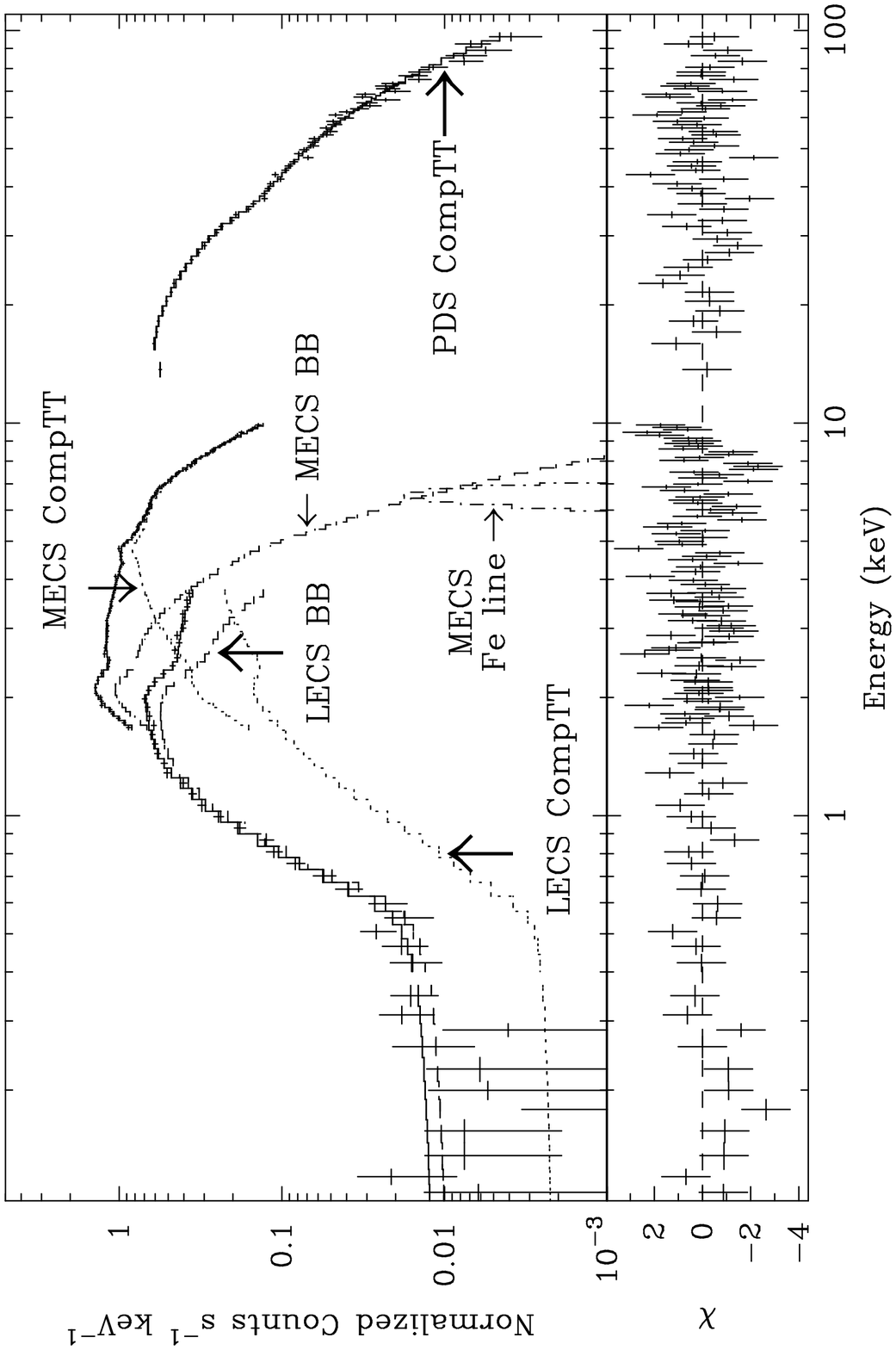}
\caption{Energy spectra of \gro\ obtained with the LECS, MECS and PDS detectors
of 2001 March 16 (left panel), April 01 (middle panel), and April 13 (right 
panel) \sax\ observations, along with the best-fit model comprising a 
blackbody component, Comptonized continuum model and a narrow iron line 
emission. The data in 0.1--4.0 keV, 1.65--10.0 keV, and 15.0--100.0 keV ranges
are used for the  LECS, MECS, and PDS detectors respectively. The individual
spectral components are also shown in the top panels. The bottom panels show 
the contributions of the residuals to the $\chi^2$ for each energy bin for 
each observations.}
\label{spec}
\end{figure*}

\section{Spectral Analysis}
\subsection{Pulse phase averaged spectroscopy}

For the spectral analysis, we have extracted LECS spectra from regions of 
radius $6 \arcmin$ centered on the object (the object was at the center of 
the field of view of the instrument). The combined MECS source photons 
(MECS~2+3) were extracted from circular regions with a $4 \arcmin$ radius. 
The response matrices released by \sax\ Science Data Center (SDC) in 1998 
November were used for the spectral fitting. Background spectra for both 
LECS and MECS instruments were extracted from the appropriate blank-fields 
with annular regions around the source. We rebinned the LECS spectra to 
allow the use of the $\chi^2$-statistic. Events were selected in the energy 
ranges 0.1--4.0 keV for the LECS, 1.65--10.0 keV for the MECS and 15.0--100.0 
keV for the PDS where the instrument responses are well determined. Combined 
spectra from the LECS, MECS and PDS detectors, after appropriate background 
subtraction, were fitted simultaneously. All the spectral parameters, other 
than the relative normalization, were tied for all three detectors.

Simultaneous spectral fitting of the broad-band energy spectrum (0.1--100 
keV) of \gro\ with a single power-law model and the line of sight absorption 
yielded a very poor fit. Addition of an iron emission line at $\sim$ 6.7 keV
and a high energy cutoff to the model, while improving the spectral fit 
slightly, produced an unacceptable reduced $\chi^2$ of $\sim$ 3. As the 
broad-band continuum spectra of a few pulsars and also the medium and hard
X-ray spectrum of \gro\ (Galloway et al. 2004) are described by a Comptonized 
component, we subsequently fitted the combined spectrum of \gro\ using a 
model consisting of a Comptonized continuum (compTT in XSPEC) with the 
geometry switch value equals to 1 (planar), along with a Gaussian function, 
and the interstellar absorption. We found that the above model requires a 
blackbody component for the soft excess to fit the broad-band energy spectrum 
of the pulsar with reduced $\chi^2$ in the range of 1.1--1.7.

In all the three \sax\ observations, the interesting results we found are :
\begin{itemize}
\item The hydrogen column density along the line of sight (N$_H$) is unusually 
low. The value of N$_H$ is found to be in the range of 4.0--5.0 $\times$ 
10$^{21}$ atoms cm$^{-2}$ whereas in the direction of the source, the 
estimated value of hydrogen column density for the entire length of the 
galaxy is $\sim$ 1.2 $\times$ 10$^{22}$ atoms cm$^{-2}$. However, using
the value of A$_v$ = 3.38 (Negueruela et al. 2003) in the relation 
N$_H$/A$_v$ = 1.79 $\times$ 10$^{21}$ atoms cm$^{-2}$ mag$^{-1}$ (Predehl \& 
Schmitt 1995), we calculate the value of N$_H$ to be 6 $\times$ 10$^{21}$ 
atoms cm$^{-2}$, in good agreement with our X-ray measurements with \sax.
\item The temperature of the soft blackbody component is found to be 
$\sim$ 0.6 keV.  The higher value of blackbody temperature for the soft
excess compared to a value of 0.1-0.2 keV of the same in several other
accreting pulsars (Paul et al. 2002) implies that the radius of the
blackbody emission region (assuming a spherical geometry and the 
distance of the source as 10 kpc) is in the range 
of 14-21 km which is very small compared to the Alfven radius. This rules 
out the inner accretion disk as the possible soft X-ray emission region. 
The contribution of the blackbody flux to the total flux is estimated to 
be 8-9\%. The blackbody radius at different pulse phases for the April 13 
observation is in the range of 11-15.5 km while the phase averaged value is 
14.5 km. The blackbody radius was calculated from the estimated blackbody 
flux by assuming a spherical emission zone. 
\item The detection of an iron emission line at $\sim$ 6.7 keV and the 
absence of the iron $K_\alpha$ emission line at 6.4 keV. The 6.7 keV iron 
line flux is found to be decreasing along with the hard X-ray flux 
during the \sax\ observations of the pulsar whereas the equivalent
width of the emission line is found to be similar (within measurement
errors). 
\end{itemize}

Among the three \sax\ observations, the source was brightest on March 16 
with a 0.1--100.0 keV flux of $\sim$ 4.3 $\times$ 10$^{-9}$ ergs cm$^{-2}$ 
s$^{-1}$ and faintest on April 13 with a flux of $\sim$ 2.0 $\times$ 
10$^{-9}$ ergs cm$^{-2}$ s$^{-1}$ in above energy band. Assuming a 
distance of 10 kpc for the source, the total broad band (0.1--100 keV) 
X-ray luminosity of the X-ray pulsar is 5.1, 3.3, and 2.4 times 10$^{37}$ 
ergs s$^{-1}$ during the three observations made with \sax\ respectively. 
Though the source intensity varies by a factor of 3, no systematic changes 
are observed in the spectral parameters. Over about 30 days, even the 
equivalent width of the iron emission has remained same within the 
measurement errors. The spectral parameters of 
the Comptonized continuum model obtained from the simultaneous spectral 
fitting to the LECS, MECS, and PDS data of the three \sax\ observations 
are given in Table~\ref{spec_par}. The count rate spectra of all observations 
are shown in Figure~\ref{spec} along with the residuals to the best-fit 
Comptonization model in the bottom panel. 

\begin{table}
\centering
\caption{Spectral parameters for \gro\ during 2001 \sax\ observations}
\begin{tabular}{llll}
\hline
\hline
Parameter          &16 March               &01 April      &13 April\\
\hline
\hline
N$_H^1$	          &4.3$\pm$0.1             &4.5$\pm$0.1            &4.3$\pm$0.3\\
$kT_{BB}$ (keV)	  &0.65$\pm$0.01           &0.62$\pm$0.01          &0.63$\pm$0.01\\
Line Energy (keV) &6.7$\pm$0.1             &6.62$^{+0.11}_{-0.06}$ &6.60$^{+0.26}_{-0.10}$\\
Line width  (keV) &0.01$_{-0.01}^{+0.13}$  &0.01$_{-0.01}^{+0.13}$ &0.01$_{-0.01}^{+0.50}$\\
Eqw. width (eV)   &30$\pm$9                &29$\pm$9               &19$^{+10}_{-12}$\\
Line Flux$^2$     &4.3$\pm$1.2             &2.7$\pm$1.0            &1.3$\pm$0.8\\
$kT_0$ (keV)	  &1.6$\pm$0.1             &1.64$^{+0.02}_{-0.01}$ &1.75$\pm$0.04\\
kT (keV)	  &12.4$\pm$0.1            &13.22$\pm$0.04         &13.7$\pm$0.5\\
$\tau$		  &2.64$^{+0.01}_{-0.03}$  &2.45$\pm$0.01          &2.33$\pm$0.11\\
Reduced $\chi^2$  &1.7 (184)               &1.2 (184)              &1.1 (178)\\
Blackbody flux$^3$   &3.8                  &2.1                    &1.6 \\
(0.1-10.0 keV range) &                     &                       &\\  
CompTT flux$^3$      &39.3                 &25.4                   &18.9 \\
(0.1-100.0 keV range) &                    &                       &  \\
\hline
\hline
\multicolumn{4}{l}{$kT_{BB}$ = Blackbody temperature, $kT_0$  = Input spectrum 
temperature}\\
\multicolumn{4}{l}{kT = plasma temperature, $^1$ : 10$^{21}$ atoms cm$^{-2}$}\\
\multicolumn{4}{l}{$^2$ : 10$^{-4}$ photons cm$^{-2}$ s$^{-1}$, $^3$ : 10$^{-10}$ ergs cm$^{-2}$ s$^{-1}$}\\
\end{tabular}
\label{spec_par}
\end{table}

\subsection{Pulse phase resolved spectroscopy}
The presence of a significant energy dependent dip in the rising part of 
the pulse profile of \gro\ prompted us to make a detailed study of the
spectral properties at different pulse phases of the neutron star.
To estimate the change in the spectral parameters at different pulse
phases, we have carried out pulse phase resolved spectroscopy during 
2001 April 13 observations in which the all three instruments had 
long exposures. The LECS and MECS spectra were accumulated into 
10 pulse phase bins by applying phase filtering in the FTOOLS task 
XSELECT. The LECS and MECS background spectra, used for phase-averaged 
spectroscopy, were used for the phase-resolved spectral fitting. 
The PDS phase resolved spectra were accumulated into 10 pulse phases 
from the corresponding event file by using the program "pdproduct". 
Appropriate LECS, MECS, and PDS response files were used for the 
phase-resolved spectral fitting. 

\begin{figure}
\vskip 8.6 cm
\includegraphics{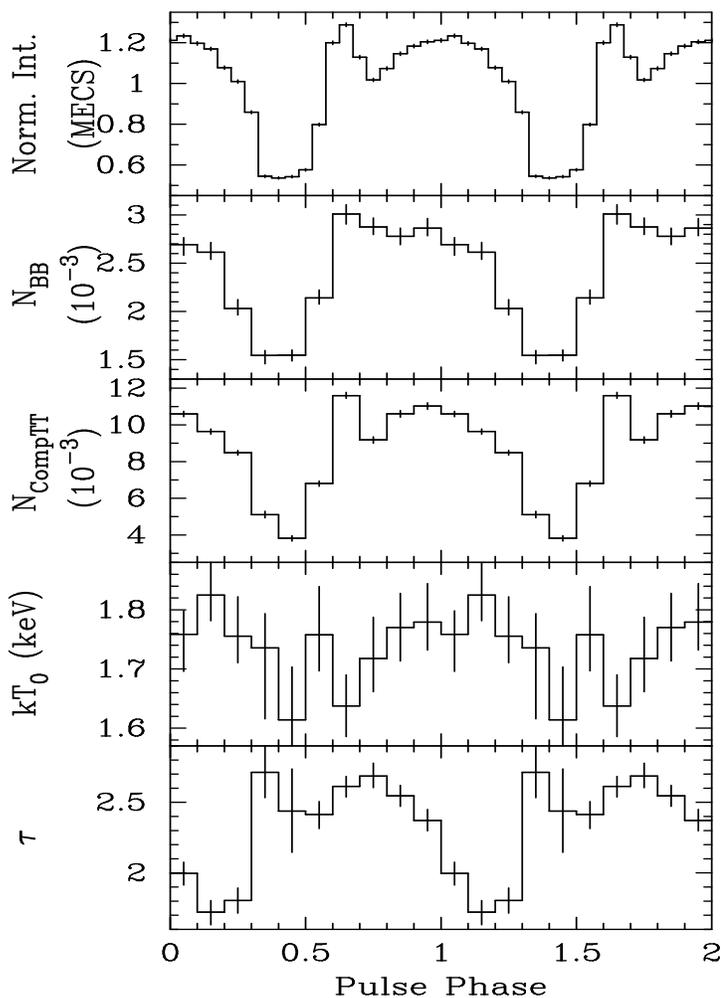}
\caption{Spectral parameters obtained from pulse phase resolved
spectroscopy of 2001 April 13 \sax\ observation of \gro. The smooth
variation of the blackbody normalization (second panel from top) indicates 
the absence of the narrow dip in soft X-rays which is clearly seen in 15-75 keV 
pulse profiles (Figure~\ref{PP}). }
\label{PRS}
\end{figure}

The phase resolved spectra were fitted with the same model used to
describe the phase averaged spectra of \gro\ keeping the iron 
emission line energy and width fixed at the phase averaged values.
The results obtained from the phase resolved spectral analysis
did not show any significant variation of N$_H$ over the pulse phase. 
This suggests that the column density is not coupled with the rotation 
of the neutron star in the binary system. The blackbody temperature 
(kT$_{BB}$) of the soft excess was also found to have similar value 
within the measurement error throughout the spin phases. As the optical
depth ($\tau$) and the temperature of the Comptonizing plasma (kT) are 
anti-correlated, we therefore froze the N$_H$, kT$_{BB}$, and kT to 
the pulse phase averaged values and repeated the pulse phase resolved 
spectroscopy. The parameters obtained from simultaneous spectral fitting 
to the LECS, MECS, and PDS phase resolved spectra are plotted in 
Figure~\ref{PRS}. The errors shown in the figure are estimated for a 90\% 
confidence level. The smooth variation of the blackbody normalization 
over the pulse phase (second panel from top) explains the absence of the 
dip feature in the pulse profiles of the soft X-ray energy bands. The 
variation of the blackbody normalization component over the pulse phase 
also indicates the pulsating nature of the soft component. The variation 
of the normalization of the Comptonized component over the pulse phase is 
shown in the third panel of Figure~\ref{PRS}. The presence of a dip like 
structure at the same phase bin to that of the pulse profile (top panel)
confirms that the dip which is absent at the soft X-rays ($<$ 3 keV)
is prominent in the hard spectral component. However, the variation of 
the other spectral components is not significant enough to get any clear 
understanding of the nature of the dip in the pulse profile.

\section{Discussion}
\subsection{Pulse period evolution and energy dependence of the Pulse profile}

The spin-down of the pulsar \gro\ was observed at an average 
rate of 8 ms yr$^{-1}$ (Swank \& Morgan 2000). \sax\ measurements 
of the spin period of KS~1947+300 on three occasions in 2001 March-April 
(present work) shows that the pulsar has spun-up rapidly within 
about 4 months. The X-ray luminosity of the pulsar during the \sax\ 
observations in 2001, when the source was in the decaying phase of 
the outburst, is in the range of 2--5 times 10$^{37}$ ergs s$^{-1}$. 
The additional torque due to the accretion of the matter which 
triggered the X-ray outburst in this transient X-ray pulsar, 
therefore has caused a transient spin-up of the pulsar. However, 
the spin evolution of the pulsar is studied in more detail from regular 
RXTE monitoring in 2000 November 21 - 2001 June 18  and 2002 March 29 - 
2002 May 17 (Galloway et al. 2004). Pulse timing analysis of above two 
sets of RXTE observations showed that the rate of increase in the pulse 
frequency is approximately proportional to the X-ray flux of the pulsar. 

During the \sax\ observations 
of \gro\ when the MECS pulse profile was characterized by a narrow peak and 
a broad peak as was seen in RXTE-PCA profiles at higher flux levels (Galloway 
et al. 2004), the energy resolved pulse profiles, as shown in Figure~\ref{PP}, 
are found to be strongly energy dependent. At low energies (0.1--3.0 keV band), 
the profile is characterized by a flat top without the narrow peak. At higher 
energies, a sharp peak followed by a dip appears at the rising part of the 
profile. This feature is seen in the pulse profiles in the energy band of 
3.0--75.0 keV beyond which the feature disappears as the X-ray light curves 
are background dominated. Pulse phase resolved spectral analysis confirmed
the absence/presence of the narrow dip in the pulse profiles because of the
absence of the notch in the blackbody component which is seen in the hard
Comptonised component. The observed change in shape of pulse profiles of
\gro\ with energy is also seen in some binary pulsars such as 4U~1626-67 
(Angelini et al. 1995; Chakrabarty et al. 1997), LMC~X-4 (Naik \& Paul 2004a), 
SMC~X-1 (Naik \& Paul 2004b), EXO~053109-6609.2 (Paul et al. 2004) etc.

\subsection{The broad band X-ray spectrum of \gro}

Since the first detection of \gro, it has been observed with the BATSE, 
RXTE, \sax, and INTEGRAL observatories. Though pulse phase averaged spectral 
studies of \gro\ have been done in the 20--75 keV energy range using X-ray 
data from the BATSE instruments (Chakrabarty et al. 1995), broad band X-ray 
spectroscopy in 0.1--100 keV energy range is reported here for the first 
time. The 20--75 keV BATSE spectra were described by a power-law with spectral 
index of 2.65 when the source was bright. However, during quiescence, the 
RXTE/PCA spectra of \gro\ above 2 keV was described by a cutoff power-law 
spectrum with a photon index of about 0.6 and e-folding energy of 10 keV 
above a cut-off energy of 6 keV with a small column density ($<$ 10$^{21}$ 
cm$^{-2}$) (Swank \& Morgan 2000). Galloway et al. (2004) also carried out a 
spectral study of the pulsar using RXTE data during the 2000-2001 and 2002 July 
outbursts. A Comptonized continuum component along with a blackbody component 
and a Gaussian representing the fluorescent Fe K$_\alpha$ emission was used to 
describe the source spectrum in 2--80 keV energy range (2--25 keV RXTE-PCA 
spectrum and 15--80 keV RXTE-HEXTE spectrum). A blackbody component with kT 
$\sim$ 3--4 keV (at the peak of the outburst) or an additional broad 
Gaussian component ($\sigma$ $\sim$ 10 keV) was also required by these 
authors to achieve the best fit to the X-ray spectra. Though the energy 
range used for spectral fitting is 2--80 keV, Galloway et al. (2004) 
estimated the value of the column densities of neutral matter to be 
consistent with the line-of-sight value i.e. $\sim$ 10$^{22}$ atoms 
cm$^{-2}$. In the hard X-ray band, the best spectral measurements 
reported so far are with the INTEGRAL, and the X-ray spectrum in the
5-90 keV band is fitted well with an exponentially cut-off power law
with cut-off energy at 8.6 keV (Tsygankov \& Lutovinov 2005).

These are the only occasions when a study of the X-ray properties 
of \gro\ has been attempted. Though a power-law continuum component was 
used to describe the spectrum of \gro\ in some of the earlier spectral 
studies in relatively narrow energy bands, it is found to be unsuitable 
when fitted to the broad-band \sax\ spectra. A Comptonization 
continuum component, as used to describe the spectrum of a few other accretion 
powered X-ray pulsars, along with a high energy cutoff and an iron emission 
line is found to provide the best broad-band spectral fit. Even though 
Galloway et al. (2004) fitted the energy spectra of \gro\ using a Comptonized 
component, the significant differences in results obtained from 2.0--80.0 keV 
RXTE spectral fitting and broad-band spectral fitting in 0.1--100.0 keV energy 
range of \sax\ data are 
\begin{itemize}
\item the temperature of the blackbody component used to explain the soft 
excess is much less ($\sim$ 0.6 keV) compared to that of 3--4 keV which was 
required to fit the RXTE spectra (Galloway et al. 2004).
\item the value of the column density measured from \sax\ data is found to 
be significantly lower than that estimated from the RXTE observations, and 
\item the iron emission line is found to be located at $\sim$ 6.7 keV in 
0.1--100.0 keV \sax\ spectra although it was reported earlier to be iron 
$K_\alpha$ emission line from the RXTE observations. 
\end{itemize}

The value of the hydrogen column density estimated from the broad-band spectral 
fitting of the three \sax\ observations is found to $\sim$ 4--5 $\times$ 
10$^{21}$ atoms cm$^{-2}$, which is about one half of the integrated value 
of the Galactic column density in this direction. This indicates a very low 
absorption by material close to the X-ray binary. Optical observations of the 
counterpart of \gro\ also suggest that the binary system is located in an 
area of low interstellar absorption slightly above the Galactic plane 
(Negueruela et al. 2003). 

It has been proposed that all/most of the accretion powered X-ray pulsars 
have an excess soft X-ray emission, the origin of which varies from source 
to source mainly due to the intrinsic X-ray luminosity (Hickox et al. 2004). 
Sources located in the Galactic plane are usually subjected to strong X-ray 
absorption and the soft excess is usually not detectable. The soft component 
is detectable only in those pulsars which do not suffer from strong absorption 
by material along the line of sight, such as LMC~X-4 (N$_H$ $\sim$ 6 $\times$ 
10$^{20}$ atoms cm$^{-2}$ : Naik \& Paul 2004a, Paul et al. 2002), SMC~X-1 
(N$_H$ $\sim$ 2--5 $\times$ 10$^{21}$ atoms cm$^{-2}$; Naik \& Paul 2004b, 
Paul et al. 2002), EXO~053109-6609.2 (N$_H$ $\sim$ 0.6--3.0 $\times$ 10$^{21}$ 
atoms cm$^{-2}$; Paul et al. 2004), Her~X-1 (N$_H$ $\sim$ 4 $\times$ 10$^{21}$ 
atoms cm$^{-2}$; Endo et al. 2002), 4U~1626-67 (N$_H$ $\sim$ 1 $\times$ 
10$^{21}$ atoms cm$^{-2}$; Orlandini et al. 1998), Cen~X-3 (N$_H$ $\sim$ 2 
$\times$ 10$^{22}$ atoms cm$^{-2}$; Burderi et al. 2000), XTE J0111.2--7317 
(N$_H$ $\sim$ 2 $\times$ 10$^{21}$ atoms cm$^{-2}$; Yokogawa et al. 2000) etc. 
The column density measurement for these X-ray pulsars have been made by 
observatories with good sensitivity at low energies, such as \sax\ and ASCA. 

In the high luminosity systems, the 
soft excess emission is likely to be from the inner part of the accretion 
disk with a temperature of about 0.1-0.2 keV. In the quiescent states of 
some accretion-powered X-ray pulsars, a higher temperature (kT$\sim$1.0 keV) 
blackbody component has been detected (Mukherjee \& Paul (2005) and 
references therein). A blackbody component with a temperature of about
0.5 keV is quite common in the anomalous X-ray pulsars (Paul et al. 2000).
However, the area of the emission zone in the two later cases are about a 
few km$^2$ and are, therefore, likely to be from the surface of the
neutron star. The soft excess emission zone in KS~1947+300 has a size of 
more than 10 km$^2$ and the accretion column is a more likely origin. 
The pulsation detected in the blackbody component also suggests the
accretion column origin of the soft excess. However, we note that Hickox 
et al. (2004) have argued against a accretion column origin of the soft 
excess from the point of view of brightness temperature.

%\subsection{6.7 keV iron emission line}
Broad-band phase-averaged spectroscopy of \gro\ shows the presence of a 
weak and narrow 6.7 keV iron emission line with equivalent width in the range
of 55--85 eV. It is interesting to note that the 6.7 keV iron line is present 
in all three \sax\ spectra of \gro\ while the 6.4 keV emission line is absent. 
The 6.7 keV line is identified as emission from helium like iron. This emission 
line might be regarded as due to the radiative recombinations of H-like iron 
followed by cascade processes in a relatively hot corona (Hirano et al. 1987). 
Similar iron emission at  6.67 keV with an equivalent width of 107 eV was 
detected in the absence of the 6.4 keV emission line in the \sax\ spectrum of 
the high mass X-ray binary pulsar Cen~X-3 (Burderi et al. 2000). The $Ginga$ 
observation of Cen~X-3 also showed the presence of 6.7 keV line during  
mid-eclipse, when the 6.4 keV line was relatively weak (Nagase 
et al. 1992). This line was interpreted as the emission from He-like iron 
ions in a hot plasma emitting resonance and fluorescence lines at 6.63, 6.67 
and 6.70 keV. By taking into account the heating of an X-ray irradiated 
plasma by Compton scattering and photoionization and cooling by thermal 
bremsstrahlung and line emission, Hirano et al. (1987) showed that the 
emission lines with energies $\geq$ 6.63 keV are emitted by recombination 
and subsequent cascades, whereas the lines with energies $\leq$ 6.63 keV 
are due to X-ray fluorescence. 

As the MECS detectors of \sax\ do not have sufficient spectral resolution 
to identify the exact ionization state of iron by separating the lines at 
energies 6.63 keV, 6.67 keV and 6.70 keV as seen in LMXBs, X-ray observations 
of \gro\ with instruments having better spectral resolution (Chandra) 
are required to investigate the presence of such lines in the energy spectrum. 
These observations may also explain the absence/presence of the soft excess 
over the Comptonized continuum model at the peak of the outburst. The 
absence of any cyclotron resonance absorption feature in the phase averaged 
spectra of \gro\ implies the cyclotron line lies outside the 0.1-100.0 keV
energy band. This suggests the magnetic field of the pulsar is very high
($\geq$ 10$^{13}$ G).

\section*{Acknowledgments}
The authors would like to thank an anonymous referee for suggestions that 
helped to correct certain issues in the paper. The \sax\ satellite was a 
joint Italian and Dutch program. We thank the staff members of \sax\ Science 
Data Center and RXTE/ASM group for making the data public. The work of SN is 
partially supported by the Japan Society for the Promotion of Science and 
IRCSET through EMBARK fellowships. 

{}

\end{document}